\documentclass[acus]{JAC2000}
\usepackage{graphicx}
\begin{document}

\thispagestyle{empty}

\onecolumn

\begin{flushright}
{\large
SLAC--PUB--8481\\
June 2000\\}
\end{flushright}

\vspace{.8cm}

\begin{center}

{\LARGE\bf
Upgrade of the PEP-II Low Beta Optics\footnote
{\normalsize{Work supported by
Department of Energy contract  DE--AC03--76SF00515.}}}

\vspace{1cm}

\large{
Y.~Nosochkov, M.~Biagini, Y.~Cai, F.-J.~Decker, M.H.R.~Donald, S.~Ecklund\\
A.~Fisher, W.~Kozanecki, P.~Krejcik, A.~Kulikov, M.~Minty, I.~Reichel\\
J.~Seeman, M.~Sullivan, U.~Wienands, Y.~Yan\\
Stanford Linear Accelerator Center, Stanford University,
Stanford, CA 94309}

\end{center}

\vfill

\begin{center}
{\LARGE\bf
Abstract }
\end{center}

\begin{quote}
\large{
The successful commissioning and operation of the PEP-II asymmetric
$e^{+}e^{-}$
collider motivated further studies to increase luminosity. In this paper,
we discuss a modification of the PEP-II lattice to reduce the vertical beta
function at the Interaction Point (IP) from the design value of
$\beta_y^*\!=\!1.5cm$ to $1.0cm$. This could potentially reduce
the colliding beam size, increase the particle density at IP and the
probability of beam-beam interactions. In this paper,
we outline the optics modifications, discuss tracking simulations, and
overview machine implementation.
}
\end{quote}

\vfill

\begin{center}
\large{
{\it Presented at the 7th European Particle Accelerator Conference 
(EPAC 2000)\\
Vienna, Austria, June 26--30, 2000} \\
}
\end{center}

\newpage

\pagenumbering{arabic}
\pagestyle{plain}

\twocolumn

\title{UPGRADE OF THE PEP-II LOW BETA OPTICS\thanks{Work supported by
Department of Energy contract DE--AC03--76SF00515.}}

\author{
\normalsize{Y.~Nosochkov, M.~Biagini, Y.~Cai, F.-J.~Decker,
M.H.R.~Donald, S.~Ecklund, A.~Fisher, W.~Kozanecki}\\
\normalsize{P.~Krejcik, A.~Kulikov, M.~Minty, I.~Reichel,
J.~Seeman, M.~Sullivan, U.~Wienands, Y.~Yan, SLAC, CA 94309, USA}}

\maketitle

\begin{abstract} 
The successful commissioning and operation of the PEP-II asymmetric 
$e^{+}e^{-}$
collider motivated further studies to increase luminosity. In this paper, 
we discuss a modification of the PEP-II lattice to reduce the vertical beta 
function at the Interaction Point (IP) from the design value of 
$\beta_y^*\!=\!1.5cm$ to $1.0cm$. This could potentially reduce
the colliding beam size, increase the particle density at IP and the 
probability of beam-beam interactions. In this paper,
we outline the optics modifications, discuss tracking simulations, and
overview machine implementation.
\end{abstract}

\vspace{-1mm}
\section{INTRODUCTION}

The luminosity of the PEP-II asymmetric $e^{+}e^{-}$ collider~\cite{cdr} has 
been steadily increased since the beams were first brought into collision
in July 1998. The present record luminosity is 
$2.2\!\cdot\!10^{33}cm^{-2}s^{-1}$ which is 73\% of the design
value. The luminosity is currently limited by the
total beam current which is below design value due to vacuum chamber
heating problems near IP in the High Energy Ring (HER) and electron cloud 
instability in the Low Energy Ring (LER). While it may
take some time before all design beam parameters are achieved, other
options can be explored to increase the luminosity. Specifically, 
the vertical $\beta_y^*$ function at the IP could be reduced below 
its design and operating value of $1.5cm$. With preserved beam emittance, 
this would result in a smaller colliding beam size, increased probability 
of particle interactions and, hence, the luminosity. 
In this paper, we discuss an upgrade of the PEP-II Interaction Region 
(IR)~\cite{optics} to reduce the $\beta_y^*$ to 
$1cm$. The potential improvement of the luminosity is up to 22.5\% if
the emittance and other beam parameters are not changed, and up to 50\%
if vertical emittance is reduced with the same rate as $\beta_y^*$.
One limitation on the minimum
$\beta_y^*$ is placed by the bunch length which is currently about 
$\sigma_{l}\!\approx\!1cm$. Reduction of $\beta_y^*$ below this level 
would not be very effective because the colliding area at the bunch ends 
($\pm\sigma_{l}$ from IP) will start to grow and offset the luminosity gain 
near the bunch center.

\vspace{-1mm}
\section{BETA SQUEEZE}

The PEP-II has been operated for more than a year
with the design values of $\beta_x^*/\beta_y^*\!=\!50/1.5cm$.
In order to raise the luminosity, our present goal is to reduce
$\beta_y^*$ to $1cm$, the level of the bunch length.
The Interaction Region has locally matched optics and
local correction systems~\cite{optics}.  
One of the requirements for the IR with lower $\beta_y^*$ is to
maintain locally matched optics and correction systems to avoid 
global optics perturbations.
This requires simultaneous adjustment of the following IR magnets:
1) quadrupoles;
2) local sextupoles compensating non-linear chromaticity 
from the final focus doublets;
3) local skew quadrupoles compensating $x$-$y$ coupling caused by the 
detector solenoid; and
4) local dipole correctors compensating orbit from the tilted solenoid.

The theoretically matched IR optics with $\beta_y^*\!=\!1cm$ was
developed for the HER and LER using MAD code~\cite{mad}.
The most noticeable and unavoidable effect caused by the lower 
$\beta_y^*$ is an enlargement of $\beta_y$ peak at the final focus 
doublets, proportional to $1/\beta_y^*$. The higher
$\beta_y$ peak further increases
sensitivity of the doublets to field, alignment and energy errors. 
At $\beta_y^*\!=\!1cm$, the natural vertical chromaticity grows by 
24\% and 15\% in the HER and LER, respectively, mainly due to increased 
contribution from the doublets. Compensation of doublet chromaticity was 
done mostly by raising the field in the vertically correcting IR
sextupoles by 37\% in HER and 22\% in LER. In the LER, 
the desired field at these sextupoles (SCY3) could not be reached 
due to magnet field limitation, hence the $\beta_y$ function at the SCY3 
was raised by $11\%$ to compensate the lack of the field. The strengths 
of the IR local skew quadrupoles were adjusted to account for the change
in optical transformation between the solenoid and the skew quads.

Because of independent optics adjustment on the left and right side of IR,
about 40 magnet families in each ring change strength to make $1cm$
lattice. The typical quadrupole field change is a few 
percent, but several quads require 10-18\% change. As stated earlier,
the sextupoles correcting the doublets require
the most raise in magnet field. The strength change of IR skew quadrupoles
have rather large variation, though many of the quads have reduced
field. At $\beta_y^*\!=\!1cm$, the new IR strengths are still
within magnet limitations except the LER SCY3 
sextupoles described above and LER SK5 and SK5L skew quadrupoles.
In the latter case the required skew quadrupole strength was made up by
creating $\sim\!2mm$ vertical orbit bumps in the SCY3 sextupoles located
next to SK5 and SK5L. The B1 and QD1 magnets near IP, shared by the two 
beams, were not changed, therefore each ring could be adjusted 
independently.

Machine implementation and tuning of the more sensitive $1cm$ optics could 
be difficult without smooth transition from the present IR configuration. 
For that reason, an intermediate matched optics with 
$\beta_y^*\!=\!1.25cm$ was designed. Furthermore, a two step linear 
``low beta knob'' was made to provide a continuous transition from
$\beta_y^*\!=\!1.5cm$ to $1.25cm$ (step 1) 
and from $1.25cm$ to $1cm$ (step 2). In each step, all IR variable 
strengths are linearly changed with
$\beta_y^*$ between two matched configurations. 
With this knob, the IR optics would not be exactly matched everywhere
except the above three $\beta_y^*$ points. However, the residual optics 
effects for entire transition are rather small. MAD calculations show
that distortions are below 0.0007 for betatron tune, 
0.2 for chromaticity and $\pm3\%$ for $\beta$ functions. Therefore 
potentially, the knob could be used not only for transition, but for 
operation at transition $\beta_y^*$ as well.
 
\vspace{-1mm}
\section{TRACKING SIMULATIONS}

At lower $\beta_y^*$, the higher sensitivity of IR doublets to
errors and stronger sextupoles could increase the effects of betatron
resonances and lead to reduced dynamic aperture, especially in the vertical
plane where most optics changes have occurred. To evaluate the impact of
the new optics on dynamic aperture, we performed tracking simulations
at the present and lower $\beta_y^*$ values, using LEGO code~\cite{lego}.
To identify the nearby resonances which may be affecting the beam
lifetime, dynamic aperture was also scanned around machine working point.

\vspace{-1mm}
\subsection{Dynamic Aperture}

The typical tracking simulation with LEGO included: 1) an assignment of 
field, multipole and alignment errors to magnets according to
PEP-II specifications; 2) global correction of tune, linear chromaticity,
coupling and orbit; and 3) tracking of particles injected at various
$x$ and $y$ amplitudes to determine dynamic aperture, 
the area of particle stable motion. 
The particles were tracked for 1024 turns with synchrotron
oscillations and initial relative energy error of $8\sigma_\delta$, 
where $\sigma_\delta$ is 0.061\% in HER and 0.077\% in LER. 
The tracking was done at the present machine tune
of $\nu_x/\nu_y=$ 24.569/23.639 in HER and 38.649/36.564 in LER, and the 
dynamic aperture was evaluated at the PEP-II injection point. The typical 
rms orbit observed in the PEP-II operation is on the order of $1mm$. For 
realistic results, the orbit correction in the simulations was adjusted 
to provide similar residual orbit. The linear chromaticity 
was corrected to zero for the tracking. The beam-beam effects were not 
included at this time.

\begin{figure}[tb]
\centering
\includegraphics*[width=82mm]{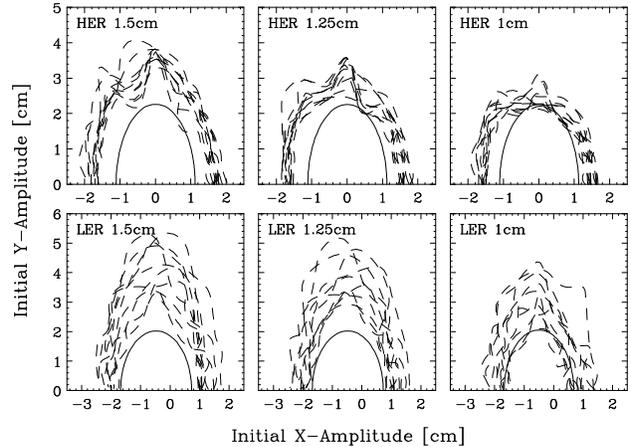}
\vspace{-5mm}
\caption{Dynamic aperture vs. $\beta_y^*$ for 10 error settings.}
\label{aper}
\vspace{-3mm}
\end{figure}

The tracking results at $\beta_y^*\!=\!1.5cm$, $1.25cm$ and $1cm$ for 
both rings are shown in Fig.~\ref{aper}, where the dash lines
correspond to dynamic aperture at injection point for 10 different 
machine error settings, and the solid ellipse shows for comparison the 
size of $10\sigma_{x,y}$ fully coupled beam with 
emittance of $\epsilon_{x}\!=\!48nm$ and $\epsilon_{y}\!=\!\epsilon_{x}/2$.
Note that normally the circulating beam with corrected coupling
has $\epsilon_{y}/\epsilon_{x}$ ratio on the order of 3\%
and the $10\sigma_y$ size four times smaller compared to the
ellipse in Fig.~\ref{aper}. However, it is important to maintain large 
vertical dynamic aperture because of the vertical injection 
with initial amplitude equal to $\sim\!8\sigma_y$ of a
fully coupled beam. The $x$-offset of the LER dynamic aperture
in Fig.~\ref{aper} is due to non-zero dispersion at the LER injection
point and $8\sigma_\delta$ initial energy error in tracking.

As shown in Fig.~\ref{aper}, the lower $\beta_y^*$ causes gradual 
reduction of 
dynamic aperture. Compared to $1.5cm$ lattice, dynamic aperture at 
$1cm$ is reduced by about 15\% and 30\% in the $x$ and $y$
planes, respectively. The vertical aperture is more affected since
most optics changes occurred in the vertical plane. For instance,
at lower $\beta_y^*$, particles in the vertical plane experience 
stronger non-linear field in the IR doublets and IR vertically correcting 
sextupoles due to larger oscillations in the doublets and stronger sextupoles.

According to Fig.~\ref{aper}, it is expected that operation at lower
$\beta_y^*$ would require more careful machine 
tuning to minimize any effects causing large particle oscillations.
It appears that injection conditions are adequate at $1.25cm$, but
become tighter at $1cm$. Note that dynamic aperture in Fig.~\ref{aper}
is for particles in the tail of energy distribution 
($\delta\!=\!8\sigma_\delta$). The particles in the beam core
will have larger aperture.

\vspace{-1mm}
\subsection{Tune Scan}

The working point used in the current PEP-II operation and in this
study differs from the design~\cite{cdr}. It was selected 
experimentally in the machine operation based on maximum luminosity, 
while the design tune was better optimized for maximum
dynamic aperture with single beam. For better understanding the
tune space and effects of betatron resonances near present working point, 
dynamic aperture tune scan was performed. In this study, 
the betatron tune was varied in 0.0025 steps around the machine working point 
within the range of $\pm0.04$ for $\nu_x$ and $\nu_y$, and dynamic aperture 
was calculated at each point. Due to extensive computing time 
in this study, the number of particle launching conditions was 
limited to five, namely with 1-2) $\pm x$, $y\!=\!0$; 
3) $x\!=\!0$, $y\!>\!0$; and 4-5) $x\!=\!\pm y$, $y\!>\!0$ initial 
amplitudes. The minimum aperture among the five conditions
at each tune point was then used to determine dynamic aperture dependence
on $\nu_x$ and $\nu_y$. As in the previous study, machine errors were
applied to the magnets, and the initial particle energy
error was set to $8\sigma_\delta$ with synchrotron oscillations included. 
Similar to machine operations, the tune 
variation was done with the ``tune knob'' which
uses pre-calculated linear dependence of quadrupole strengths in the tune
sections with the tune.

\begin{figure}[t]
\centering
\includegraphics*[width=65mm]{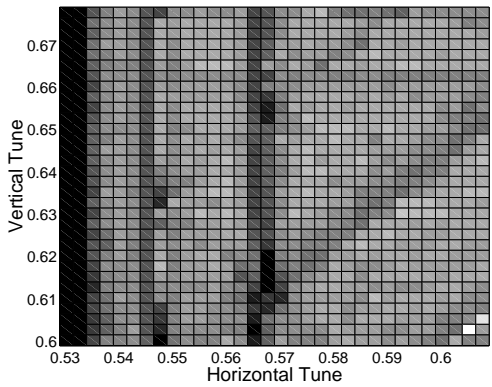}
\vspace{-1mm}
\caption{HER tune scan for $\beta_y^*=1cm$.}
\label{herscan}
\vspace{-1mm}
\end{figure}

\begin{figure}[t]
\centering
\includegraphics*[width=65mm]{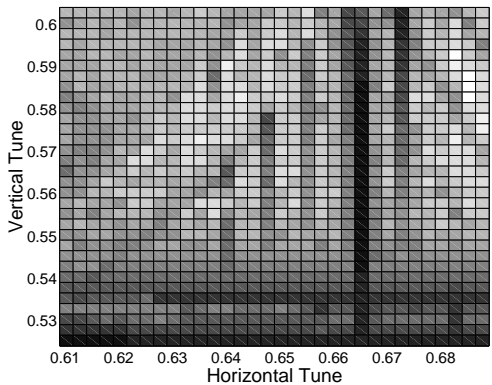}
\vspace{-1mm}
\caption{LER tune scan for $\beta_y^*=1cm$.}
\label{lerscan}
\vspace{-2mm}
\end{figure}

The HER and LER tune scan diagrams for $\beta_y^*\!=\!1cm$ are shown in
Fig.~\ref{herscan} and \ref{lerscan}, where a smaller aperture
corresponds to a darker shade and the present working point is in the center.
Numerical analysis of the tune scan reveals several lines of reduced dynamic
aperture on the tune plane, associated with synchro-betatron 
resonances. 

In the HER, the strongest resonance identified near working point is
$2\nu_x\!+\!k\nu_s\!=\!49$, where $k\!=\,$-1,-2,-3 and the synchrotron 
tune $\nu_s\!=\!0.044$. The dynamic aperture practically disappears 
at the first synchrotron side band of the $2\nu_x$ 
resonance. Two other weaker resonances were observed: 
$3\nu_y\!=\!71$ and $\nu_x\!-\!\nu_y\!+\!\nu_s\!=\!1$. 
According to the tune scan, the HER aperture
could be improved by moving further away from the nearby
$2\nu_x\!-\!3\nu_s$ resonance.

In the LER, the following resonances were identified:
1) $2\nu_y\!+\!l\nu_s\!=\!73$, with $l\!=\,$-2,-3; 
2) $3\nu_x\!+\!m\nu_s\!=\!116$, with $m\!=\,$-2,-1,0,1; and 
3) $2\nu_x\!+\!\nu_y\!+\!n\nu_s\!=\!114$, with $n\!=\,$1,2,3 
and $\nu_s\!=\!0.025$. 
The first two resonances above are the strongest and result in
rather small or vanished dynamic aperture close to resonance 
conditions. The effect is more pronounced near the lowest order 
synchrotron side bands of the resonances. 
Compared to $1.5cm$ lattice, at $\beta_y^*\!=\!1cm$ 
the $2\nu_y\!+\!l\nu_s$ resonance is enhanced due to increased 
vertical chromaticity and sensitivity in the IR doublets, 
thus further limiting the available vertical tune 
space near the working point. According to the tune scan, the LER dynamic 
aperture could be improved by increasing the $\nu_y$ and reducing the
$\nu_x$ tunes by $\sim\!0.015$, however the beam-beam conditions 
in this area have not been verified yet.

Note that the tune scan results are valid for one particular set of
machine errors and particles with $\delta\!=\!8\sigma_\delta$ oscillations.
Particles with smaller energy error will be less affected by the 
resonance synchrotron side bands.

\vspace{-1mm}
\section{MACHINE IMPLEMENTATION}

The first step of the described low beta modification has been recently 
implemented in the machine, and the IR optics is currently set at 
$\beta_y^*\!=\!1.25cm$. The transition to $1.25cm$ lattice was done
using the ``low beta multi-knob'' described earlier, with the beams stored 
in the machine. In this way, any residual distortions (tune shifts, 
chromaticity, etc.) could be compensated as they appeared in the 
transition, and any effect of the larger beam size in the IR quadrupoles 
on the backgrounds could be detected immediately.
The spurious tune shift observed during this operation was about 
0.006 in $x$ and $<\!0.001$ in $y$ planes. The global change
in the beam orbit was $\sim\!0.4mm$, which was easily compensated with orbit 
correctors. Residual chromaticity amounted to less than one unit 
in either plane.

\begin{figure}[htb]
\centering
\includegraphics[width=75mm]{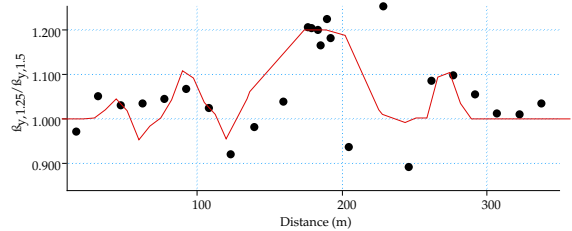}
\vspace{-1mm}
\caption{Ratio of HER $\beta_y$ functions near IP at
$\beta_y^*\!=\!1.25cm$ with respect to $\beta_y$ values at
$\beta_y^*\!=\!1.5cm$.}
\label{bratio}
\vspace{-0mm}
\end{figure}

The effect of the new $\beta_y^*$ on lattice functions was verified 
using the on-line phase advance and $\beta$ function measurement facility.
Fig.~\ref{bratio} shows the ratio of the new HER $\beta_y$ functions 
with respect to the old values, measured at BPMs 
(circles) near IP (at center of the figure). The agreement with the 
MAD prediction (solid line) is satisfactory.

No increase in the background was observed during and after implementing
the $1.25cm$ lattice. Luminosity did not visibly increase immediately 
after the $\beta_y^*$ change. A few days later, solenoids to
reduce the effect of photoelectrons on the LER beam were powered up.
After that, luminosity increased significantly, raising the record
from the previous value of $1.6\!\cdot\!10^{33}$ to 
$1.95\!\cdot\!10^{33}cm^{-2}s^{-1}$.

\end{document}